\documentclass[showpacs, oneside, twocolumn, prl, amsmath, amssymb, nofootinbib, superscriptaddress]{revtex4-2}

\usepackage{cases}
\usepackage{amsmath}
\usepackage{amssymb}
\usepackage{amsfonts}
\usepackage{amssymb}
\usepackage{dcolumn}
\usepackage{bm}
\usepackage{bbm}
\usepackage{graphicx}
\usepackage{xcolor}
\usepackage{array}
\usepackage{subfigure}
\usepackage{hyperref}
\usepackage{wasysym}
\usepackage{booktabs}

\newcommand{\be}{\begin{equation}}
\newcommand{\ee}{\end{equation}}
\newcommand{\ba}{\begin{eqnarray}}
\newcommand{\ea}{\end{eqnarray}}

\newcommand{\gsim}{\mathrel{\hbox{\rlap{\lower.55ex \hbox {$\sim$}}
                   \kern-.3em \raise.4ex \hbox{$>$}}}}
\newcommand{\lsim}{\mathrel{\hbox{\rlap{\lower.55ex \hbox {$\sim$}}
                   \kern-.3em \raise.4ex \hbox{$<$}}}}

\renewcommand{\(}{\left(}
\renewcommand{\)}{\right)}
\renewcommand{\[}{\left[}
\renewcommand{\]}{\right]}

\def\be{\begin{equation}}
\def\ee{\end{equation}}
\newcommand{\bea}{\begin{eqnarray}}
\newcommand{\eea}{\end{eqnarray}}
\def\fnl{f_{\rm NL}}

\hypersetup{colorlinks=true,
            breaklinks=true,
            pdfstartview=Fit,
            linkcolor=blue,
            citecolor=blue,
            urlcolor=blue}

\begin{document}
\title{How Significant are Cosmological Collider Signals in the Planck Data?}
 
\author{Petar Suman}
\affiliation{Centre for Theoretical Cosmology, Department of Applied Mathematics and Theoretical Physics,
University of Cambridge,
Wilberforce Road, Cambridge, CB3 0WA, U.K.}

\author{Dong-Gang Wang}
\affiliation{Institute for Advanced Study and Department of Physics,\\ Hong Kong University of Science and Technology, Clear Water Bay, HK, China}
\affiliation{Centre for Theoretical Cosmology, Department of Applied Mathematics and Theoretical Physics,
University of Cambridge,
Wilberforce Road, Cambridge, CB3 0WA, U.K.}

\author{Wuhyun Sohn}
\affiliation{AstroParticule et Cosmologie 
10, rue Alice Domon et Léonie Duquet, 75013 Paris, France}

\author{James Fergusson}
\affiliation{Centre for Theoretical Cosmology, Department of Applied Mathematics and Theoretical Physics,
University of Cambridge,
Wilberforce Road, Cambridge, CB3 0WA, U.K.}

\author{E. P. S. Shellard}
\affiliation{Centre for Theoretical Cosmology, Department of Applied Mathematics and Theoretical Physics,
University of Cambridge,
Wilberforce Road, Cambridge, CB3 0WA, U.K.}

\begin{abstract}

The search for primordial non-Gaussianities (PNG) is theoretically well motivated but remains observationally challenging. 
Tight constraints with low significance for the standard non-Gaussian shapes suggest that detection may lie beyond the reach of near-future experiments. However, tests of PNG are highly template-dependent. 
From a theory perspective, a whole new family of bispectrum shapes arise in the cosmological collider program, with distinct signatures of heavy particles during inflation. 
In this work, we provide a class of simplified collider templates for these particles that encompasses a broader range of masses, sound speeds, and interactions.
We propose that, given the current state of observations, the most effective strategy to search for PNG signals is through orthogonalizing the collider templates, such that they are uncorrelated with the tightly constrained single field predictions.
Our results show that removing the single-field contribution enhances the signal across all four templates studied.
Using the {\it Modal} pipeline and \textit{Planck} CMB data, we perform a systematic parameter scan of the (\textit{orthogonalized}) collider templates with the most significant result reaching $2.4\sigma$ for spin-0, after taking into account the look-elsewhere effect; indicative results for spin-1 and spin-2 peak around 2$\sigma$. These results indicate that, with refined collider templates and improved data analysis strategies, there are credible prospects with forthcoming observations to detect PNG {\it and also} identify the presence of massive particles during inflation.
\end{abstract}

\maketitle

{\it Introduction}--
In modern cosmology, there is an increasing recognition that the primordial Universe can serve as a natural laboratory for probing fundamental physics at extremely high energies. 
In particular, the Hubble scale of cosmic inflation has $H\lesssim 10^{13}$GeV, which is far above the upper limit of terrestrial collider experiments in the foreseeable future.
Physical processes during inflation may leave imprints in today's cosmological observations, such as surveys of the Cosmic Microwave Background (CMB) and Large Scale Structure (LSS).
Therefore, noting that inflation could be the highest-energy event accessible in nature, signatures of new physics from this primordial epoch might offer the ultimate hope of discovering the fundamental microphysical laws or even unveiling key mysteries about quantum gravity \cite{Meerburg:2019qqi,Achucarro:2022qrl}.

A major observational window for probing new physics during inflation is primordial non-Gaussianity (PNG).
Capturing deviations from Gaussian statistics of primordial fluctuations, the PNG at leading order is depicted by the scalar bispectra in Fourier space, with amplitude parameter $\fnl$. 
Importantly, PNG is {\it not} fully described by a single number. 
As the bispectra are functions of three Fourier modes ${\bf k}_1,{\bf k}_2,{\bf k}_3$, there can be many possibilities for the momenta dependence, also known as {\it shapes}, which reflect distinct predictions from various theories of inflation (see, e.g., the review \cite{Chen:2010xka}).
As a consequence, observational tests of PNG are highly shape-dependent, and using different bispectrum templates in the data fitting will yield distinct observational constraints. Therefore, for the detection of PNG, it is very important to identify various bispectrum shapes from well-motivated theories and prepare the corresponding templates.
 
Among many theory proposals, the ``cosmological collider" is particularly attractive and promising.
It draws a close analogy with the conventional particle collider, where model-independently the characteristic signatures of new heavy particles are imprinted in the statistics of primordial fluctuations \cite{Chen:2009zp,Baumann:2011nk,Noumi:2012vr,Arkani-Hamed:2015bza}. 
In the bispectrum, one typical collider signal is the oscillations in the squeezed limit for particles with mass $m>3H/2$, and the spin of the particle manifests itself as the characteristic angular dependence in the shape function.
Thus, through observations, we will be able to decode the masses and spins of these unknown high-energy particles.
This has been an active area of research in the past several years. 
At the theory frontier, new analytic tools, including the cosmological bootstrap, have been developed to compute the full shape functions of the cosmological collider (see, for example, Refs. \cite{Arkani-Hamed:2018kmz, Baumann:2022jpr, Pimentel:2022fsc, Jazayeri:2022kjy, Qin:2022fbv, Wang:2022eop, Qin:2025xct} and references therein). 

For the detection of PNG, most efforts so far have been focused on the three simplest predictions: local, equilateral and orthogonal templates. 
The \textit{Planck} experiment placed tight constraints on their amplitude parameters \cite{Akrami:2019izv}, though the Modal \cite{Fergusson:2008ra,Fergusson:2009nv}, generalised KSW \cite{Meerburg:2010osc,Munchmeyer2014kswosc}, and binned \cite{Bucher_2010} bispectrum estimators allowed for a broader survey. While having $\fnl=0$ is still fully consistent for the simplest shapes, new shapes that are {\it uncorrelated} with them may provide statistically significant levels of PNG.
Recently, observational tests of the cosmological collider signals were performed using both the CMB \cite{Sohn:2024se} and the LSS data \cite{Cabass:2024wob}. In particular, the maximum signal-to-noise ratio identified using the {\it Planck} data is $1.8\sigma$, after taking into account the look-elsewhere effect.
Although no significant signal has been detected, these preliminary attempts have paved the way toward a systematic search for the presence of unknown particles during inflation.
Meanwhile, at the current stage of observation with a low signal-to-noise ratio, it is important to ask what would be the best strategy for testing PNG.

In this Letter, we present an update on the CMB test of the cosmological collider signals that makes full use of the currently available Planck data. More details are presented in a companion paper \cite{Suman:2025tpv}.
Compared to our previous work, the improvements can be seen in three aspects: i) we include a more complete menu of the collider templates with various masses, spins and sound speeds; ii) to minimize the overlaps with the single-field equilateral-type shapes and extract the unique signatures of massive particles, we {\it orthogonalized} the collider templates; iii) {we perform the data analysis with {\it Modal}, a principal bispectrum pipeline used in the Planck analysis and independent from our previous analysis framework.}
With these efforts, our aim is to identify the following: 

$\bullet$\hspace{4pt}the most significant collider signals in the latest data;

$\bullet$ the most efficient data analysis strategies for testing them at the current stage.

{\it A complete set of scalar bispectra}--
Let's first prepare the analytical templates for bispectrum shape functions. 
For the theory setup, we assume that inflation can be described by a scale-invariant and weakly coupled effective field theory (EFT) with the inflaton field $\phi$ and additional massive particles $\sigma$ \cite{Cheung:2007st,Lee:2016vti,Bordin:2018pca,Pimentel:2022fsc,Jazayeri:2022kjy}.
Then at the leading order, the dimensionless shape function of a generic scalar bispectrum takes the following form:
\be \label{shape}
S(k_1,k_2,k_3) = x S^{\rm equil} + y S^{\rm ortho} + \sum_{n} z_n S_{\rm s.e.}^{(n)}~,
\ee
where the linear combination of $S^{\rm equil}$ and $S^{\rm ortho}$ captures the PNG from self-interaction of the inflaton fluctuations. The presence of the massive particles leads to the single-exchange shapes, and we use the sum over $S_{\rm s.e.}^{(n)}$ to represent possible contributions from various masses, spins, sound speeds and interactions.

The next step is to derive analytical templates for the massive-exchange correlators. 
Thanks to recent developments on the cosmological bootstrap, we can now not only determine these shapes precisely for any kinematics, but also provide a systematic classification of all possible predictions based on minimal assumptions. In general, for a given massive-exchange process, there are two types of contributions in the primordial bispectrum.
First, the EFT background $S_{\rm bkg}$ which consists of rational polynomials of the kinematic variables and resembles the equilateral or orthogonal templates. 
Secondly, the non-analytic collider signal $S_{\rm col.}$ which exhibits squeezed-limit oscillations. 
The latter provides the characteristic imprint of the new massive particles that cannot be mimicked by any single field shapes. 
There can be various ways of separating the collider template from the background.
Our choice is to take the squeezed limit of the full shape and then simply extend it to non-squeezed configurations of kinematics.\footnote{See \cite{Suman:2025tpv} for discussions about how this decomposition can be achieved in the exact analytical expressions of full shapes. Also see \cite{Qin:2025xct} for other proposals of the collider templates.}
Schematically, for a given mass, spin, interaction and sound speed ratio, $S_{\rm col.}$ takes the following form
\ba \label{Scol}
S_{\rm col.} &= &P_s(\hat{\bf k}_2\cdot \hat{\bf k}_3) \(\frac{k_3}{k_1+k_2}\)^{1/2}\sum_a f_a(k_i;\mu)  \\
&&~~~~~\times\cos\[\mu \log\(\frac{k_3}{2c_s(k_1+k_2)}\) +\delta_a\] +{\rm perm.}~,\nonumber
\ea
where $\mu \equiv \sqrt{m^2/H^2-9/4}$ is the mass parameter, $s$ is the spin, and $c_s\in (0,\infty)$ is the ratio between the inflaton and $\sigma$ field sound speeds.
The form factors $f_a$, the phase parameters $\delta_a$ and the sum over $a$ depend on the form of the interaction. 

With this knowledge, we may absorb the $S_{\rm bkg}$ parts of massive-exchange shapes into the two single field PNG templates in \eqref{shape}. Then a general bispectrum template can be rewritten as 
\be \label{shape2}
S(k_1,k_2,k_3) = \tilde{x} S^{\rm equil} + \tilde{y} S^{\rm ortho} + \sum_{n} z_n S_{\rm col.}^{(n)}~.
\ee
The advantage of this decomposition is that, we can separate out the truly unique signatures of massive particles during inflation.
Here we choose to focus on the massive exchanges with lower-derivative interactions and lower spins, but we leave the masses and sound speed ratios as free parameters.
Table \ref{tab:templates} summarizes the notations for collider templates that we use in the data analysis, and their corresponding spin and interactions.
We leave their analytical expressions in the companion paper \cite{Suman:2025tpv}.
These templates provide a well-motivated set of observational targets for the search of the collider signals.

\begin{table}[h]
    \centering
    \renewcommand{\arraystretch}{1.5} 
    \begin{tabular}{cccc}
    \toprule
    template &    ~~spin~~ &      cubic mixing~ & ~quadratic mixing \\
    \midrule
       $S_{\rm col.}^{\rm I}$ & 0 & $\dot\phi^2\sigma$   & $\dot\phi\sigma$ \\
        $S_{\rm col.}^{\rm II}$  & 0 &  $(\partial_i\phi)^2\sigma$ & $\dot\phi\sigma$ \\
      $S_{\rm col.}^{s=1}$ & 1 & $\dot\phi\partial_i\phi\sigma_i$  & $\partial_i\phi\sigma_i$  \\
      $S_{\rm col.}^{s=2}$ & 2 & $\dot\phi\partial_{ij}\phi\sigma_{ij}$ & $\partial_{ij}\phi\sigma_{ij}$
         \\
    \bottomrule
    \end{tabular}
    \caption{Collider templates with the corresponding intermediate states and interactions. For each template, the mass $\mu$ and sound speed ratio $c_s$ are the two free parameters, while the phases $\delta_a$ and form factors $f_a$ are fixed by $\mu$ and $c_s$. }
    \label{tab:templates}
\end{table}

{\it Orthogonalizing collider templates}--
Before moving to the CMB data analysis, let's take a closer look at the degeneracy present in the observational tests of cosmological collider. 
Quantitatively, we can compute the similarity between two shape functions as ${ {\rm cos}( S^{(1)},S^{(2)}) } \equiv {\langle S^{(1)},  S^{(2)}\rangle}/{\sqrt{\langle S^{(1)},  S^{(1)}\rangle \langle S^{(2)},  S^{(2)}\rangle}}$, where 
the inner product is defined as
\be \label{overlap}
\langle S^{(1)},  S^{(2)}\rangle \equiv \iiint_{\mathcal{V}_k} w(k_i) S^{(1)}(k_i) S^{(2)} (k_i)  \mathrm{d}\mathcal{V}_k~,
\ee
with weight $w\equiv {1}/{(k_1+k_2+k_3)}$. This weight function is chosen  to reflect the $l-$scaling of the $f_{\rm NL}$ estimator in the CMB. 
It has been noticed that in general the single-exchange shapes in \eqref{shape} may have large correlations with the two single field templates \cite{Sohn:2024se}.
If this is the case, then the significance of cosmological collider-like PNG signals depends on the parameters governing the single-field self-interactions. The observational constraints place bounds on a combined PNG from the collider and single-field, not individually.

To break the degeneracy, one can adopt the approach used in \cite{Cabass:2024wob}, where the authors included the spin-0 massive-exchange templates in \eqref{shape} and then \textit{marginalized} over all possible values of the amplitudes $x$ and $y$ of the equilateral and orthogonal shapes.
{Through marginalization, one effectively derives the bounds on $\fnl$s of collider signals assuming the model \eqref{shape}, with no prior information on the single-field interactions.} Note that this effectively places bounds on the amplitude of the \textit{part} of the collider shape that is orthogonal to the equilateral and orthogonal shapes. For a simplified example $S'_\mathrm{col.}=S_\mathrm{col.}+\alpha S^\mathrm{equil}$, this approach leads to identical marginalized constraints for $\fnl^\mathrm{col.}$ regardless of $\alpha$.

Instead, we propose the {\it orthogonalization approach} to directly measure the characteristic imprints of massive particles during inflation. The idea is to subtract the equilateral and orthogonal components from the template \eqref{shape2} directly such that the remaining shape function is uncorrelated with both of them.
In spirit, this is similar to the proposal of the original orthogonal template for single field inflation in \cite{Senatore:2009gt}. 
In practice, we construct the orthogonalized template for the collider shape as
\be \label{Sorcol}
\tilde{S}_{\rm col.} = S_{\rm col.} + \alpha  S^{\rm equil} + \beta S^{\rm ortho} ,
\ee
where we impose the following conditions to fix the two coefficients $\alpha$ and $\beta$
\be
 \langle \tilde{S}_{\rm col.}, S^{\rm equil} \rangle =0~~~{\rm and}~~~ \langle \tilde{S}_{\rm col.}, S^{\rm ortho} \rangle =0~.
\ee
The inner product is defined in \eqref{overlap}. Afterwards, we directly constrain the $\fnl$ corresponding to $\tilde{S}_\mathrm{col.}$.

This approach has several advantages. First, the resulting constraints are more easily interpretable. The $\fnl$ corresponds directly to the PNG of the shape $\tilde{S}_\mathrm{col.}$. Second, the full set of templates designed in this approach can be easily applied in any observational dataset, and thus constraints from different surveys can be compared on an equal footing. 
Third, for the linear bispectrum model in the CMB context, the constraints obtained from orthogonalized templates are easier to compute, while remaining equivalent to the marginalized constraints up to a normalization factor.\footnote{The equivalence here holds as long as the inner product defined in \eqref{overlap} yields correlations between shapes similar to those defined in the late-time CMB space. This assumption does not always hold exactly: two shapes that are orthogonal in the primordial $k$-space can exhibit correlations of $\mathcal{O}(10\%)$ at the $\fnl$ level when evaluated using the CMB bispectrum estimator. For our purposes, however, this difference is not large enough to affect the statistical significance of the results presented.}
Last, this approach highlights the characteristic signatures of the cosmological collider, as the resulting template {\it cannot} be mimicked by any single field predictions.
If the signal-to-noise ratio becomes high enough for one of these new collider templates, we detect PNG for the first time, and even more excitingly, we rule out single field inflation.

{\it Modal analysis}--
PNG leaves distinct imprints through linear effects on the CMB bispectrum. Computing the CMB bispectrum from the primordial shape requires evaluating
a four-dimensional integral over the momenta $k_i$ and the {line-of-sight distance} $r$. In addition to the prohibitively high computational cost scaling as $\mathcal{O}(l_{\rm max}^5)$ of the CMB bispectrum, another numerical challenge of these integrals comes from the mixing of kinematic variables. These difficulties can be overcome by approximating the bispectrum as a sum of separable functions, or \textit{modes}. The \textit{Modal} estimator uses two such expansions, one for the primordial bispectrum in $k-$space, and another for the CMB bispectrum in $l-$space. In this way, the computationally expensive task of CMB bispectrum estimation is efficiently performed using separable modal basis functions. The formalism and its code implementation have been rigorously validated and applied extensively in the \textit{Planck} analysis \cite{Planck:2013wtn,Planck:2015zfm,Akrami:2019izv}. In this work, we utilise the same pipeline to test the new templates of cosmological collider signals.

Note that our previous analysis \cite{Sohn:2024se} implements the CMB-BEST pipeline \cite{Sohn:2023fte}. 
This public code uses separable modal basis functions to expand a given template, but differs from {\it Modal} in that it does {\it not} have a second set of basis functions in the late-time bispectrum space. Rather, a single primordial set of basis functions is used to represent any template in a separable form, which is then constrained using a KSW-like estimator. Because of this simplification, CMB-BEST allows us to easily test arbitrary primordial bispectrum shapes on a local computer. 

The Modal and CMB-BEST pipelines have been internally cross-validated for both single field shapes \cite{Sohn:2023fte} and multiple collider templates \cite{Suman:2025tpv}. We find that the results from the two are consistent for all templates where the modal expansion remains accurate. In addition to Modal's historical legacy of robust constraints, this pipeline has proven to be faster and more efficient in constraining a large multi-dimensional parameter space, such as the example here with $\mu$ and $c_s$ covering multiple orders of magnitude. Ultimately, we aim to achieve a consistent set of constraints from multiple and mutually independent CMB bispectrum estimators. Any potentially large signal detections by both pipelines would represent a significant result to the science community. Providing first-ever {\it Modal} constraints on this new class of templates is a crucial step towards this goal.

{\it Results}--
We provide $f_{\rm NL}$ constraints for the four classes of cosmological collider templates in Table \ref{tab:templates} using the official Planck \texttt{SMICA} $T+E$ CMB map. For each class, we consider both the collider template $S_{\rm col.}$ in \eqref{Scol} and the fully orthogonalized template $\tilde S_{\rm col.}$ in \eqref{Sorcol}.
All templates are normalized at each $\mu$ and $c_s$ to yield unity in the equilateral limit: $S(k,k,k) = 1$, following the convention of the {\it Planck} analysis. This choice is useful for comparing the size of the bounds ($\sigma(\fnl)$) between different shapes. Note, however, that the signal-to-noise ratio (SNR) does not depend on this normalization convention.

\begin{figure}
    \centering
    \includegraphics[width=1\linewidth]{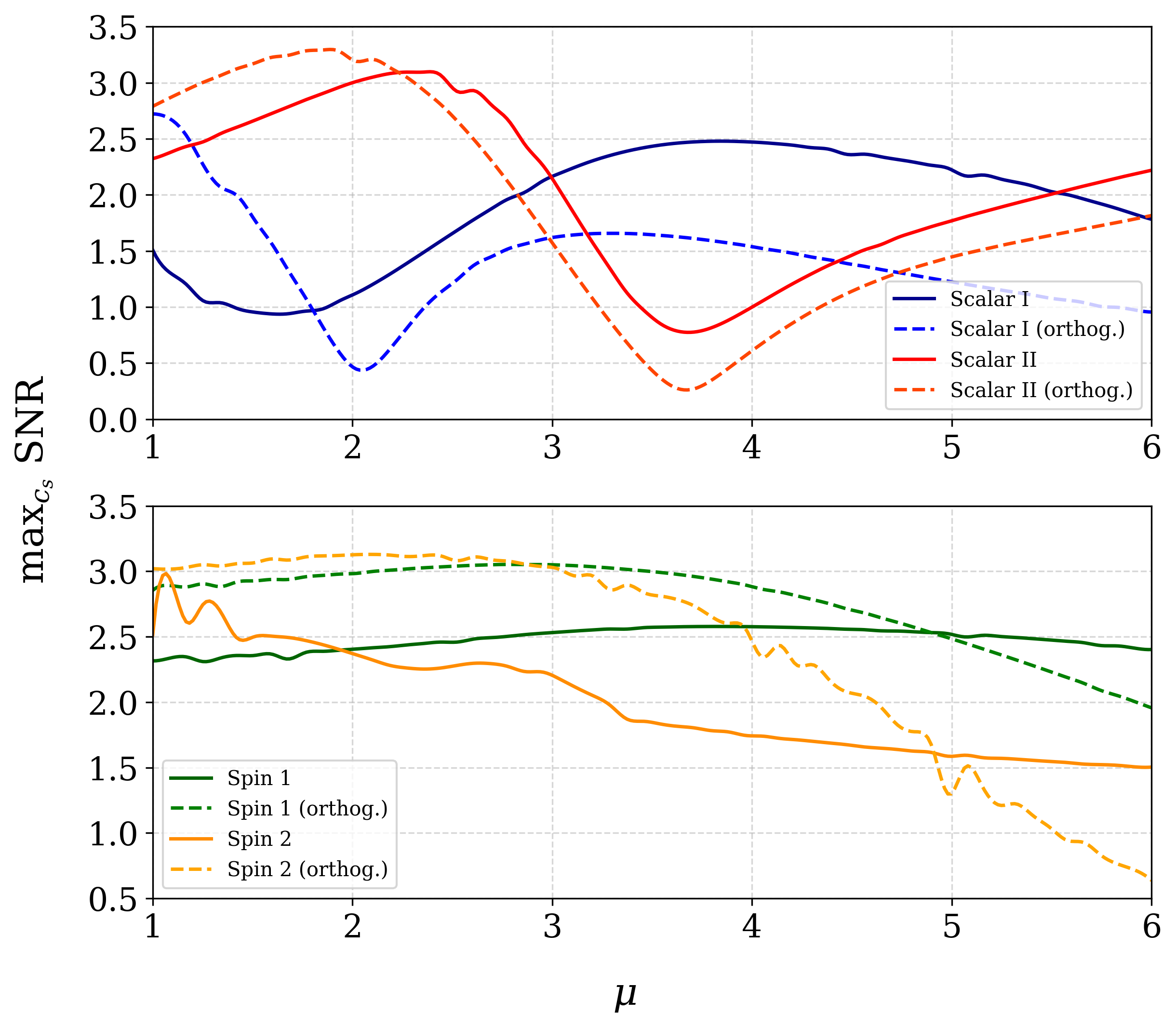}    \caption{Raw signal-to-noise ratios of the best-fit $\fnl$ as a function of the mass parameter $\mu$, for all four templates investigated in this paper: Scalar-I, Scalar-II, Spin-1 and Spin-2, and their orthogonalized counterparts. Before taking into account the look-elsewhere-effect, the most significant signals correspond to orthogonalized Scalar-II ($3.29\sigma$) and orthogonalized Spin-2 ($3.13\sigma$).}
    \label{fig:snr_plots_condensed}
\end{figure}

We perform a vast parameter search in the log-uniform range of $c_s$ between $10^{-2}$ and $10^2$, and in the uniform mass range of $\mu$ between $1$ and $6$. Any masses heavier than this would yield highly oscillating (and exponentially suppressed) shapes for which the accuracy of \textit{Modal} decomposition drops low. Raw SNR maximized over $c_s$ as a function of mass parameter is shown in Figure \ref{fig:snr_plots_condensed}. We also summarize the best-fit parameters along with highest SNR after adjusting for the look-elsewhere-effect in Table \ref{tab:constraint}.

\begin{table}[ht]
    \centering
    \begin{tabular}{c c c c}
    \hline
    \hline
    \vspace{2pt}
         \textbf{Shapes} & 
         \textbf{$f_{\rm NL}$} & \textbf{Adjusted} &
         \textbf{Best-Fit} \\
         & 
         ~~\textbf{($68\%$ CL)}~~ & \textbf{SNR} &
         ~~\textbf{Parameters}~~ \\
    \hline
    \hline
        \vspace{3pt}
    $S_{\rm col.}^{\rm I}$ &   $182\pm 73$   & $1.53$ & $\mu=3.88$ , $c_s=45.82$ \\
    $\tilde{S}_{\rm col.}^{\rm I}$ &   $3566\pm 1310$   & $1.71$ & $\mu \lesssim 1.00$ , $c_s=0.12$ \\ 
     \hline   \vspace{3pt}
     $S_{\rm col.}^{\rm II}$ &  $-295\pm 95$   & $2.14$ & $\mu=2.36$ , $c_s=0.12$\\
     $\tilde{S}_{\rm col.}^{\rm II}$ &   $-600\pm 182$  & $2.35$ & $\mu = 1.85$ , $c_s = 0.09$ \\ 
        \hline   \vspace{3pt}
        $S_{\rm col.}^{s=1}$   & $-132\pm 51$   & $1.86$ &   $\mu=3.80$ , $c_s=2.76$\\
        $\tilde{S}_{\rm col.}^{s=1}$  &  $483\pm 158$   & $2.14$ &  $\mu = 2.78$ , $c_s=1.73$ \\ 
                \hline   \vspace{3pt}
   $S_{\rm col.}^{s=2}$ & 
   $19\pm 6.5$  & $1.93$ &   $\mu=1.09$ , $c_s=0.68$ \\
       $\tilde{S}_{\rm col.}^{s=2}$ &  $70\pm 22$  & $1.98$ &  $\mu=2.10$ , $c_s=0.07$ \\
                                \hline
    \end{tabular}
    \caption{Summary of CMB constraints on cosmological collider templates and the best-fit values of the mass parameter $\mu$ and $c_s$, the sound speed ratio between the inflaton and $\sigma$ field. {The $\fnl$ constraints stated are evaluated at the best-fit parameters of $(\mu,c_s)$.}}
    \label{tab:constraint}
\end{table}
As we can see, {\it Planck CMB} data favours different values of the mass parameter $\mu$ for the two scalar exchange templates: Scalar-I shows the highest signal-to-noise at intermediate masses around $3\lesssim \mu \lesssim 5$, while Scalar-II has its best fit 
around $2 \lesssim \mu \lesssim 3$. Both best-fit signals are enhanced after orthogonalization, suggesting that the bispectrum contributions from single-field interactions were partially responsible for the weaker signal. This is in line with Planck constraints on single-field shapes which are consistent with vanishing PNG, especially in the case of equilateral template. Favoured values of $\mu$ shift to $\sim 1$ for Scalar-I and $\sim 2$ for Scalar-II after orthogonalization. 
For the Spin-1 template, the SNR remains almost constant for the whole range of the mass parameter under consideration.
This is because for odd-spin particles, the oscillatory collider signals are more suppressed than the EFT background, and thus the overall shape becomes less dependent on the mass parameter \cite{Lee:2016vti,Pimentel:2022fsc}.
For the Spin-2 template, the highest SNR  is observed for small masses $\mu\simeq1$, which then decreases as we increase the mass.
Again, the SNR becomes higher in general when we include the orthogonalized templates into the analysis.

\begin{figure}
    \centering
\includegraphics[width=0.47\textwidth]{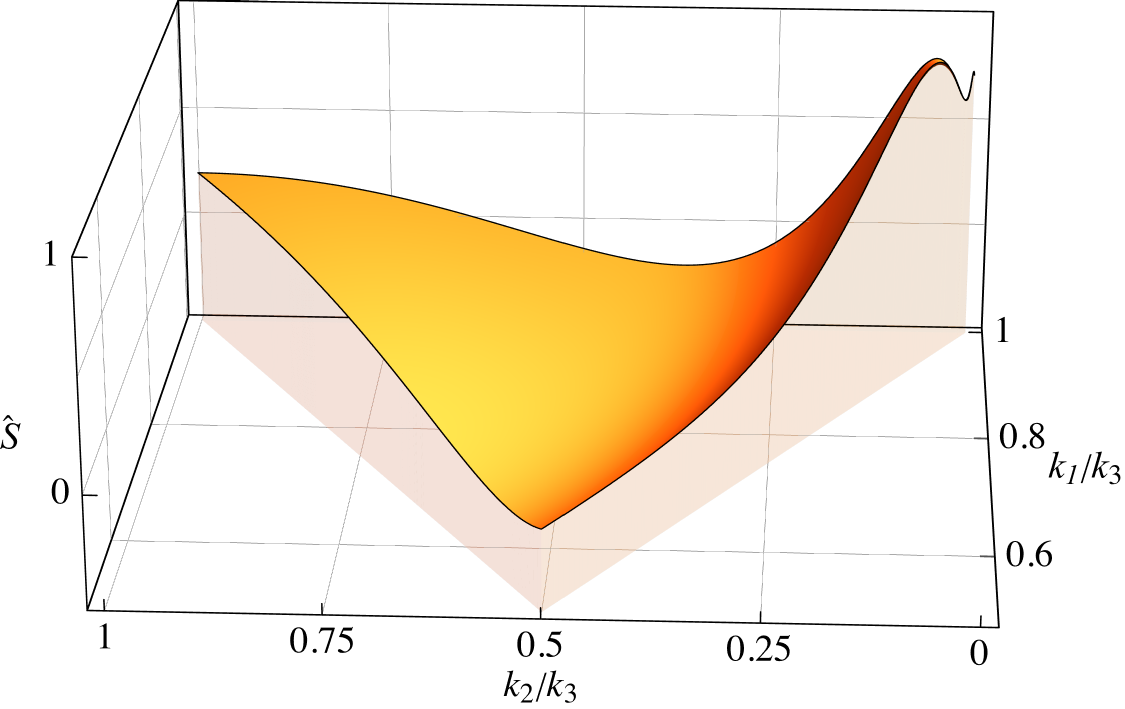}
    \caption{The most favoured primordial shape from orthogonalized Scalar-II template with $\mu_*=1.85$ and $c_{s*}=0.09$. It corresponds to $\tilde{S}^{\rm II} = S^{\rm II}_{\rm col.}(\mu_*, c_{s*}) - 0.253\, S^{\rm equi.} - 0.237\, S^{\rm ortho.}$.}
    \label{fig:best}
\end{figure}

The primordial template in our analysis that is most favoured by the Planck data is shown in Figure \ref{fig:best}. After taking into account the look-elsewhere effect, the adjusted SNR is $2.35\sigma$. This shape function originates from the massive scalar exchange bispectrum $S_{\rm col.}^{\rm II}$ with $(\partial_i\phi)^2\sigma$ and $\dot\phi\sigma$ interactions. It is the orthogonalised template which has its overlap with the two single-field EFT shapes removed, while also demonstrating the characteristic oscillatory signals of the massive scalar particle.
This is consistent with the previous result of the CMB-BEST analysis \cite{Sohn:2024se}, where the most significant signal of $1.8\sigma$ comes from a similar template with $\mu=2.13$.
We leave discussion of the corresponding CMB bispectrum shape to the companion paper \cite{Suman:2025tpv}.
Other significant signals are given by orthogonalized spin-1 and spin-2 templates; with 
$(\mu =2.78, c_s=1.73)$ and an adjusted SNR of $2.14$ for $\tilde{S}_{\rm col.}^{s=1}$; and with $(\mu =2.10, c_s=0.07)$ for $\tilde{S}_{\rm col.}^{s=2}$ with an adjusted SNR of $1.98$. 
These results might provide interesting hints for the existence of unknown particles -- one for massive scalars, another possibly for KK gravitons in the very early Universe –– but a more definitive analysis must await improved data.

{\it Discussions}--
In this Letter, we update the search for cosmological collider signals of PNG in the latest Planck CMB data \cite{Akrami:2019izv} using the Modal pipeline \cite{Fergusson:2008ra,Fergusson:2009nv}. 
Our main results and their implications for future research are summarized in the following two aspects.

On the one hand, we establish a complete and optimized strategy for probing cosmological colliders at the current stage.
First, we propose a full set of collider templates under the simple theoretical assumptions of scale invariance and weak couplings. Then,  we orthogonalize these templates to break the degeneracy with single field PNG shapes.
As we have argued, there are two different approaches, marginalization and orthogonalization, that can help us identify the observational constraints for signatures of massive particles only, independently from the single field contributions. We adopt the orthogonalization approach to obtain easily interpretable constraints that can be robustly compared between different observational datasets. Our analysis constitutes the first step in detecting PNG for cosmological colliders.

Using the Modal pipeline and a systematic scan of a large parameter space for orthogonalized collider templates, we find the most significant signal of massive particles during inflation in the Planck data. The result of $2.35\sigma$ after taking the look-elsewhere effect into account is given by an orthogonalized massive scalar exchange template with $\mu=1.85$. This may be hinting towards a non-zero scalar bispectrum in the Universe that we live in (see Figure \ref{fig:best}). 
If this is confirmed with more precise measurements in the future, it would  provide a concrete direction for developing our theoretical understanding of the physics of inflation.
In general, we find that the orthogonalized templates have higher signal-to-noise ratios than the original collider templates. This can be explained by the fact that the equilateral and orthogonal shapes have low significance in the Planck data, and thus removing them improves the ability to detect the distinctive collider signals.

While we focus on the {\it Planck} constraints on cosmological collider signals in the current work, our data-analysis strategy can be easily applied in other observational tests of PNG.
For instance, using this approach, it will be straightforward to include the new dataset from the Simons Observatory \cite{ade2019simons}, which, augmented by Planck at low multipoles, is expected to significantly tighten the constraints on the collider signals. 
It will also be interesting to extend our analysis to collider signals in the primordial trispectrum, which contains richer structures arising from spinning particles during inflation. One recent attempt in this direction is presented in \cite{Philcox:2025bvj,Philcox:2025wts}.
Meanwhile, multiple LSS experiments, such as SphereX, Euclid and LSST, will deliver a wealth of data in the next decade that can be used for testing cosmological collider signals \cite{MoradinezhadDizgah:2017szk,MoradinezhadDizgah:2018ssw,Goldstein:2025eyj}. In particular, a recent forecast study \cite{Anbajagane:2025uro} has shown that these galaxy surveys can lead to constraints on collider PNG competitive with the previous CMB results in \cite{Sohn:2024se}.
It would be very informative to find out if more accurate and more general templates here would lead to stronger constraints in the forecast. The orthogonalized collider templates developed in this work would be useful for future PNG studies with both CMB and LSS, and would provide a common platform to obtain joint constraints.

\vskip12pt
{\it Acknowledgments}-- We are grateful to Santiago Agüi Salcedo, Xingang Chen, Giovanni Cabass, Ciaran McCulloch, Xi Tong and  Bowei Zhang for stimulating discussions.
We also thank the anonymous referee for constructive comments.
JF and PS are supported by Science and Technology Facilities Council (STFC) grant ST/X508287/1. DGW is partially supported by a Rubicon Postdoctoral Fellowship from the Netherlands Organisation for Scientific Research (NWO), the EPSRC New Horizon grant EP/V048422/1, and
the Stephen Hawking Centre for Theoretical Cosmology, and GRF Grant 16306425 from the Research Grants Council of Hong Kong. WS is supported by the
SCIPOL project funded by the European Research Council (ERC) under the European Union’s Horizon 2020 research and innovation program (Grant agreement No. 101044073). Part of this work was undertaken on the Cambridge CSD3 part of the STFC DiRAC HPC Facility (www.dirac.ac.uk) funded by BEIS capital funding via STFC Capital Grants ST/P002307/1 and ST/R002452/1 and STFC Operations Grant ST/R00689X/1.

\bibliography{references}
\end{document}